\documentstyle[epsfig, twocolumn]{esapub}

\begin{document}

\setlength{\parindent}{0pt}
\setlength{\parskip}{ 10pt plus 1pt minus 1pt}
\setlength{\hoffset}{-1.5truecm}
\setlength{\textwidth}{ 17.1truecm }
\setlength{\columnsep}{1truecm }
\setlength{\columnseprule}{0pt}
\setlength{\headheight}{12pt}
\setlength{\headsep}{20pt}
\pagestyle{esapubheadings}

\title{\bf \Large ON SOLAR FREQUENCY CHANGES}

\author{{\bf M.~Cunha$^1$, M.~Br\"uggen$^1$, D.O. Gough$^{1,2}$} \vspace{2mm} \\
$^1$Institute of Astronomy, Madingley Road, Cambridge CB3 0HA, UK \\
$^2$Department of Applied Mathematics and Theoretical Physics, Silver Street, Cambridge CB3 9EW, UK}

\maketitle

\begin{abstract}
The structure of the surface layers of the Sun is changed by 
magnetic activity which, in turn, changes the eigenfrequencies of the 
acoustic modes. These frequency shifts have been observed both in low- and 
high-degree data, and are found to be correlated with the solar cycle. 

Time-distance helioseismology has shown that waves travelling
through sunspots suffer time delays. Also it is known that sunspots
induce phase shifts between inwardly and outwardly propagating
waves. 

In the present work we show how these phase shifts are related to the
observed time delays by treating the scattering of acoustic waves off
sunspots as a diffraction problem.

These time delays are then related to the frequency shifts of the
modes.  From this relation we obtain a lower limit for the
contribution of sunspots to the observed frequency variations with the
solar cycle. Moreover, we predict the frequency dependence of the time
delays suffered by rays travelling through a sunspot. \vspace {5pt} \\

Key~words: helioseismology; solar cycle; sunspots.

\end{abstract}

\section{INTRODUCTION}

Since the mid eighties (Woodard \& Noyes 1985) there has been clear
evidence that the frequencies of acoustic oscillations in the Sun
change with the solar cycle.  These frequency shifts have now been
observed over the period of a solar cycle (Elsworth {\it et al.} 1994,
Jimen{\' e}z-Reyes 1998), and frequency differences of the order of a
few hundred of nanoHertz have been found between the two extremes of
the cycle.
 
Since then, a significant amount of work has been undertaken in order
to study the dependence of these shifts on many parameters, such as
the frequency, degree and azimuthal order of the modes, as well as on
different time-scales and on the different epochs of the solar cycle.
As a result of these studies it was shown that the frequency shifts
vary with frequency and degree of the modes in a way that is
proportional to the inverse of the mode inertia (Libbrecht \& Woodard
1990), suggesting that this effect is due to a change at the surface,
rather than due to a change in the structure of the Sun's
interior. The dependence on the azimuthal order, on the other hand,
allowed for the study of the latitudinal dependence of the surface
perturbation frequency shifts (Kuhn 1988, Gough 1988, Libbrecht \&
Woodard 1990, Woodard \& Libbrecht 1993), showing that it follows the
latitudinal variations of the effective temperature. Moreover, with
another solar maximum coming up, new results are starting to appear
(Dziembowski {\it et al.} 1997) which seem to agree well with those of
the previous solar cycle.

The time-scales on which the frequency variations take place were also
the subject of several studies. In addition to the long-term
variations following the period of the solar cycle, Woodard {\it et
al.} (1991) and Bachmann \& Brown (1993) found short-term frequency
changes, taking place over periods as short as one month. More
recently, Jimen{\' e}z-Reyes (1998) also found such short-term
frequency shifts, this time using low-degree data.

The good correlation between the long-term frequency shifts and
several solar-cycle indices (e.g Bachmann \& Brown 1993, Elsworth {\it
et al.} 1994, Jimen{\' e}z-Reyes 1998) supports the idea that these
changes are due to either the direct effect of the magnetic field on
the oscillations or its indirect effect, via changes in the Sun's
thermal structure, or both. Exactly which of these effects is the most
important is still unknown, although efforts have been made to
reconcile the observed frequency shifts with the magnetic activity and
the surface brightness variations (Goldreich {\it et al.} 1991,
Balmforth {\it et al.} 1996, Kuhn 1998). Moreover, considering the
effect of the magnetic field on the frequency shifts, it is still
uncertain which magnetic regions contribute more significantly for
these shifts. Is it the quiet sun, the sunspots, or other magnetically
active regions? The theories that explore the direct effect of the
magnetic field on the frequencies commonly use a regular perturbation
method, which is valid only when the magnetic perturbations are small,
breaking down in regions such as sunspots, where the magnetic field is
strongly concentrated near the surface. As a result, the contributions
arising from such regions are difficult to assess. Knowledge about the
extent to which various regions and effects (thermal or magnetic)
influence the solar frequencies would help us to know what solar cycle
indices we should be correlating the frequency shifts with.

A possible way to study the effect of localized structure variation
could be via time-distance helioseismology. In the present work we use
the modal in combination with the time-distance approach to
helioseismology in an attempt to isolate the contribution of sunspots
to the frequency shifts of normal modes. Our eventual goal is to
relate the time delays as observed by time-distance helioseismology of
sunspots to the frequency shifts of global modes. Thus we can
establish a lower limit for frequency shifts caused by the sunspots
and predict the dependence on the frequency of the time delays
suffered by rays travelling through a region with a surface anomaly.

\section{SUNSPOT SEISMOLOGY}

Recently sunspots have been studied seismically by two completely different
methods. The first one uses a Fourier-Hankel decomposition of the
solar oscillations in the vicinity of the sunspot to calculate the
complex amplitudes of inwardly and outwardly propagating p modes
(`modal approach') (Braun, Duvall, \& LaBonte 1988; Braun, LaBonte, \&
Duvall 1990, Braun {\it et al.} 1992; Bogdan {\it et al.} 1993; Braun
1995).

{\small
 \begin{figure}[h]
   \begin{center}
   \leavevmode
   \centerline{\epsfig{file=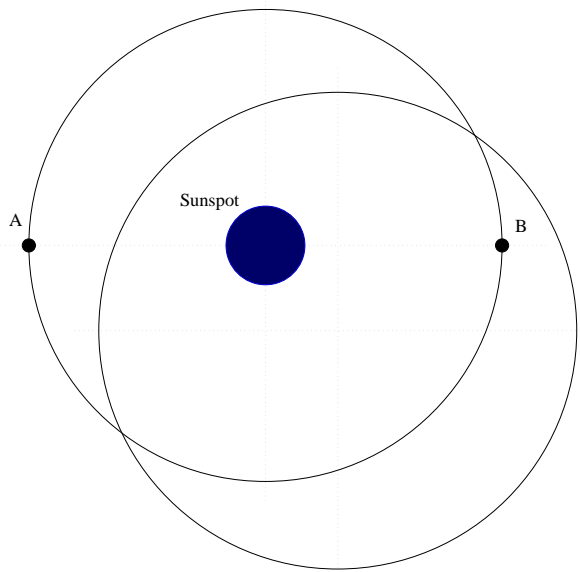,
               width=8.0cm,angle=270}}
   \end{center}
 \caption{\em Measurement of time-delays in the vicinity of sunspots.}
 \label{a}
 \end{figure}
  }

These studies generally find an attenuated amplitude of the outwardly
propagating wave and a positive phase shift relative to the inwardly
propagating wave.

The second method measures the temporal cross-correlation between
oscillations recorded at two diametrically opposite points on a circle
centred on the sunspot (see Fig.\ref{a}) (Duvall {\it et al.} 1993;
Duvall 1995; Korzennik, Noyes, \& Ziskin 1995; D'Silva, \& Duvall
1995; Duvall {\it et al.} 1996; D'Silva {\it et al.} 1996; Kosovichev
1996; Braun 1997).

{\small
 \begin{figure}[h]
   \begin{center}
   \leavevmode
   \centerline{\epsfig{file=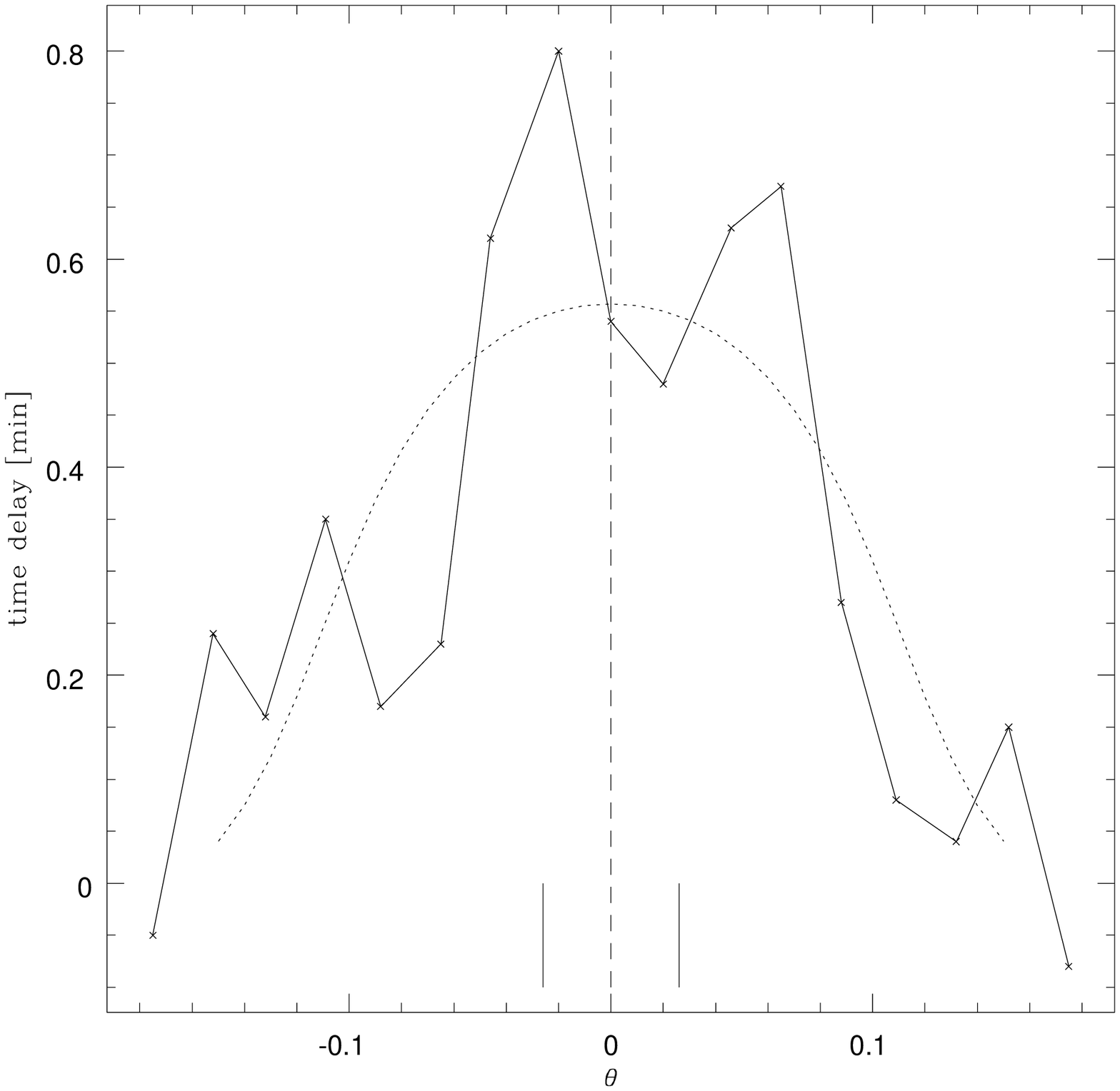,
               width=8.0cm}}
   \end{center} \caption{\em Travel time difference in the presence of
 a sunspot as compared to the quiet sun as a function of the offset
 of the origin from the spot (from Duvall 1995). The marks on the
 bottom indicate the extent of the penumbra. Superimposed (dotted line) is the
 theoretically expected time delay for those parameters shown in Fig.\ref{g}.}  \label{b} \end{figure}
  }

The signature of the sunspot manifests itself in travel-time anomalies
relative to their quiet-sun counterparts. Assuming geometrical
acoustics to be valid, Duvall (1995) deduced a phase shift of
30$^\circ$ induced by one particular sunspot (corresponding to a time
delay of 0.56 min at 3 mHz). Various models have been invoked to
explain the observed phase shifts and the attenuation of the
amplitude, including downflows beneath sunspots or conversion to slow
magnetoacoustic waves within the sunspot. However, to date, no
conclusive explanation has been found. In fact, in most cases the
interpretation of the observations is still subject to speculation.
The idea behind the work we present here is to relate these
travel-time anomalies to changes in eigenfrequencies of acoustic
modes.

\section{\sc TRAVEL-TIME DIFFERENCE VS FREQUENCY SHIFT }

\subsection{Normal modes}

We first concentrate on the analysis of normal modes, and choose a
radius $R^*$ in the region of propagation of the eigenmode which is
below the region perturbed by the sunspot but still close to the
surface (Fig. \ref{match}).

{\small
 \begin{figure}[h]
   \begin{center}
   \leavevmode
   \centerline{\epsfig{file=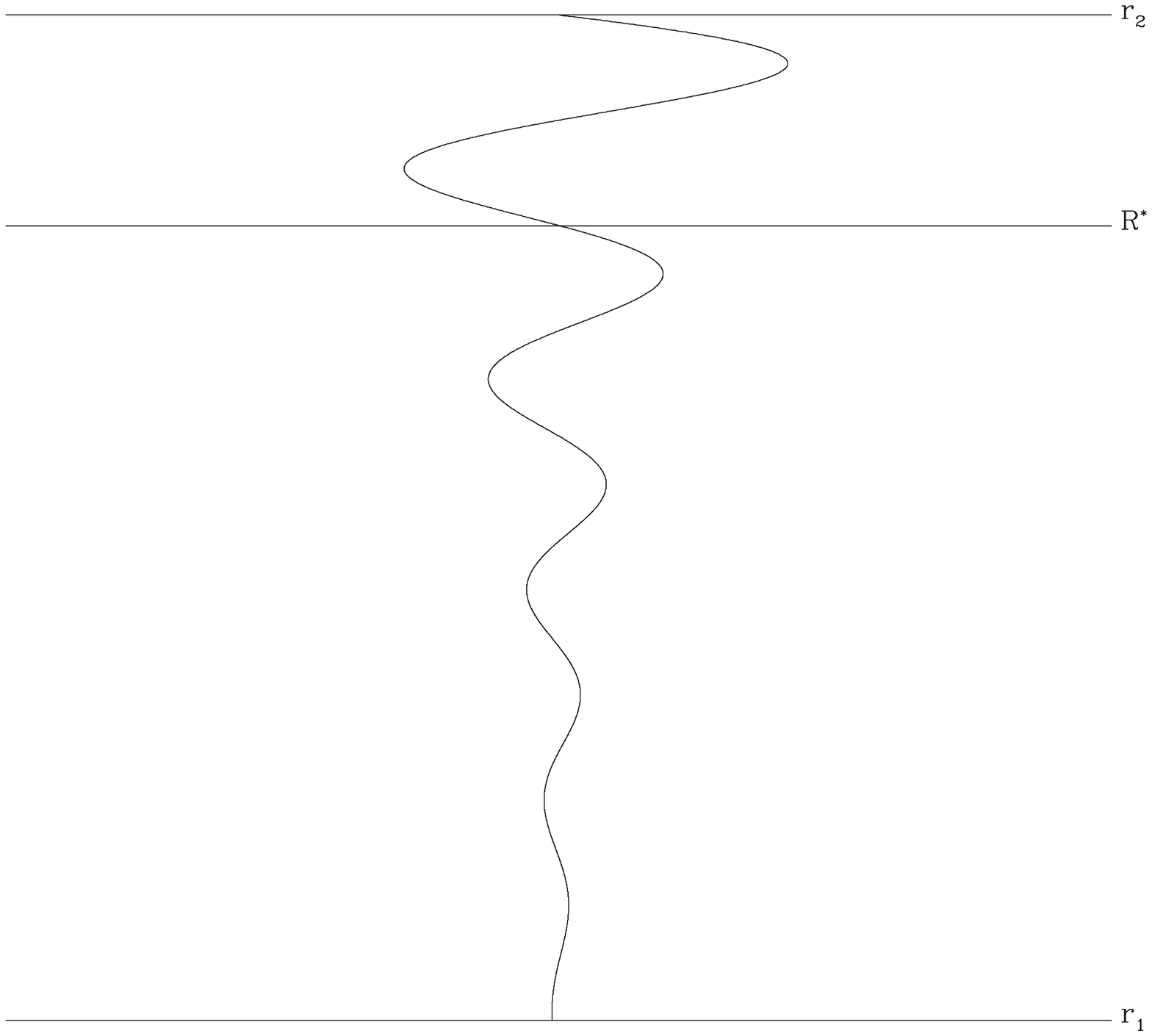,
               width=7.0cm,angle=270}}
   \end{center}
 \caption{\em Diagram of the propagating region of an eigenmode; $r_1$ and $r_2$
are the lower and upper turning points, respectively.}
 \label{match}
 \end{figure}
  } 

Note that, even if the sunspot extends considerably into the Sun's
interior, its effect is important only near the surface, due to the
rapid downward increase in the gas pressure, and, therefore, it is always
possible to choose a suitable value for $R^*$.  In the interior of the
Sun, defined here as the region between the centre and $R^*$, the
oscillations obey the usual equations of hydrodynamics.  Therefore, we
can express any perturbation above $R^*$ as a change in the boundary
condition applied to these equations at $r=R^*$.  This change in the
boundary condition at $r=R^*$ generates a shift in the
eigenfrequencies of the modes, and it is precisely this shift that we
want to obtain.  In principle, the change in the boundary condition
will not only change the eigenfrequencies of the modes but also the
eigenfunctions, which will no longer be proportional to single spherical harmonics, as
result of the latitudinal dependence of the new boundary
condition. Fortunately, however, it is possible to use a variational
principle to relate the change in the boundary condition to the
frequency shifts, while avoiding the tedious work of calculating the
perturbed eigenfunctions.  The variational principle can be expressed
as:
\begin{equation}
\frac{\Delta\omega}{\omega_0}=\frac{\overline{\pi\Delta\varphi}}{\int_{r_1}^
{R^*}\omega_0^2 c^{-2}\kappa^{-1}{\rm d}r} \label{eq:dw},
\end{equation}
where $\omega_0$ is the unperturbed frequency, $c$ is the sound speed,
$\kappa$ is the vertical wavenumber and $\pi\Delta\varphi$ is the
change in the phase of the local eigenfunction at $r=R^*$, which
constitutes the change in the boundary condition to which we referred
above. The bar over the phase shift stands for the average over
latitude, i.e.
\begin{equation}
\overline {\pi\Delta\varphi}=\int_0^{2\pi}\int_{-1}^1{\pi\Delta\varphi\left
({Y_l}^m\right)^2{\rm d}\left(cos\theta\right){\rm d}\phi}, \label{eq:avalfa}
\end{equation}
where ${Y_l}^m$ is a spherical harmonic.
So, equation (\ref{eq:dw}) can be used to convert a perturbation 
in the local phase of the eigenmode at $r=R^*$ into a shift in its
eigenfrequency.

\subsection{Travel times}

Having obtained the relation between phase shift and frequency shift,
the next step is to infer this phase shift from the travel-time
anomalies produced by the sunspots.
To do that, let us go back to the ray paths and consider again a wave
travelling between any two surface points A and B. The starting point A
can be regarded as a source point of waves, each travelling in a
different direction. Therefore, at point B the wave travelling
horizontally will interfere with the wave travelling along the ray path
joining A and B.  The condition for constructive interference is that
(Gough 1984):
\begin{equation}
\int_A^Bk_{\rm s}{\rm d}s-\int_A^Bk_{\rm h}R{\rm d}\theta+\Phi=2\pi n 
\label{eq:quantization}, 
\end{equation}
where $k_{\rm s}$ is the wavenumber along the ray path, $k_{\rm h}$ is the
horizontal wavenumber and $\Phi$ is a phase which accounts for any
phase jump that the ray may suffer on reflection.

But $k_{\rm s}=\omega /c_{\rm ph}$, where $\omega$ is the frequency of
the wave considered and $c_{\rm ph}$ is its phase velocity. Therefore,
writing the travel time of the wave as $\tau_0=\int_A^B c_{\rm ph}^{-1}{\rm
d}s$, equation (\ref{eq:quantization}) becomes:
\begin{equation}
\omega\tau_0-\int_A^Bk_{\rm h}R{\rm d}\theta+\Phi=2\pi n 
\label{eq:quantization2}.
\end{equation}
Now, from time-distance analysis we know that the travel time is
altered by the presence of a sunspot. So, if we write the new travel
time as $\tau=\tau_0+\Delta\tau$ and substitute it in equation
(\ref{eq:quantization2}), we get an extra phase jump of
$\Delta\Phi=\omega\Delta\tau$.

\subsection{From time-distance to normal modes}

The last step consists in relating the phase jump $\Delta\Phi$ to
the shift in the phase of normal modes, $\pi\Delta\varphi$.

Using the equation describing the ray paths, equation (\ref{eq:quantization2}) 
can be rewritten as:

\begin{equation}
\int_{r_1}^R \sqrt{\frac{\omega^2}{c^2}-\frac{L^2}{r^2}}=\pi n-\frac{\Phi}{2},
\end{equation}

where $L/r=k_{\rm h}$. This last equation resembles the quantization
equation for normal modes, which is usually written as:

\begin{equation}
\int_{r_1}^{r_2}\kappa{\rm d}r=\pi\left(n+{\alpha}\right), 
\label{eq:quantization4}
\end{equation} 
where $r_2$ is the upper turning point of the eigenmode, $n$ is the
radial order of the mode and ${\alpha}$ is yet another phase.
Perturbing equation (\ref{eq:quantization4}) and relating it to
equation (\ref{eq:quantization}), where $\Phi$ is perturbed due to the
sunspot, we find:
\begin{equation}
\int_{r_1}^{r_2}\Delta\kappa{\rm d}r=-\frac{\Delta\Phi}{2}=-\frac{\omega_0
\Delta
\tau}{2}, \label{eq:shif1}
\end{equation}  
where $r_2$ is the upper turning point of the normal mode.
But the phase jump $\pi\Delta\varphi$ in the normal modes is just
\begin{equation}
\pi\Delta\varphi=\int_{r_1}^{R^*}\Delta\kappa{\rm d}r=-\frac{\omega_0\Delta
\tau}{2}-\int_{R^*}^{r_2}\omega_0 c^{-2}\kappa^{-1}{\rm d}r\Delta\omega,
\label{eq:shif2}
\end{equation} 
and, therefore, substituting $\pi\Delta\varphi$ back in equation 
(\ref{eq:dw}), we obtain for $l=0$ modes:
\begin{equation}
\frac{\Delta\omega}{\omega_0}=-\frac{\omega_0\Delta\tau}{\int_{r_1}^
{r_2}\omega_0^2 c^{-2}\kappa^{-1}{\rm d}r}\frac{A_{\rm sp}}{A_{\odot}}
\label{eq:dw2},
\end{equation}
where $A_{\rm sp}/A_{\odot}$ is the ratio between the sunspot area and
the surface area of the Sun.

At this point it is probably a good idea to stop and explore equation
(\ref{eq:dw2}) more carefully. As was mentioned in section 1, it is
known from observations that the frequency shifts of global modes
follow the inverse of the mode inertia. The contributions to the
frequency shifts from the sunspot regions have a similar
dependence. Hence, we can predict the dependence of $\Delta\tau$ on
the frequency of the rays considered, by writing,
\begin{equation}
\Delta\tau=-\frac{C\Delta\omega}{I{\omega_0}^2}\int_{r_1}^{r_2}\omega_0^2 c^{-2}\kappa^{-1}{\rm d}r\frac{A_{\odot}}{A_{\rm sp}}
\label{eq:dtau},
\end{equation}  
where $C$ is a constant. We hope that in the near 
future it will be possible to test this frequency dependence of the time 
delays against observational data. 

Another point that deserves some attention is the dependence of the
frequency shifts on the degree $l$ of the modes. Aside from
the dependence on $l$ due to the mode inertia, the frequency shift
varies with the degree of the mode also due to the asphericity of the
surface perturbations. Equation (\ref{eq:dw2}) was written for the
case $l=0$, but it is straightforward to write an equivalent
expression for other degrees simply by using the definition
(\ref{eq:avalfa}) to take the average of equation (\ref{eq:shif2}),
and substituting it in equation (\ref{eq:dw}). Having done that, it is
possible to compare the outcome with frequency shifts suffered by
modes of different degrees, and, in principle, extract
information about the asphericity of the perturbations, since
perturbations at different locations in the sun affect modes of
different degrees in different ways.

In this work we use the time delays published by Duvall (1995).  The
area covered by sunspots at a given time was obtained from the
international relative sunspot number, where, the area in millionths
of the visible hemisphere is given by $A_{\rm sp}=16.7R_i$ and $R_i$
is the relative sunspot number.  Using those data we obtain for a mode
of degree $l=0$ and a frequency of 3 mHz, a frequency shift of about one
tenth of the observed frequency difference between the two extremes of the
solar cycle for a mode with the same characteristics.  We must
emphasize, however, that this value is only a lower limit to the
contribution of the sunspots to the variations in the
eigenfrequencies. In our calculation we have ignored two effects that
will increase this contribution: Firstly, in the upper layers of the
sun geometric acoustics breaks down and, consequently, we must allow
for diffraction effects when relating the observed time delay with the
phase jump $\Delta\Phi$ induced by the sunspots. As we shall see in
section 4, this phase jump is actually larger than the value deduced
from geometrical acoustics; secondly, the area cover by sunspots (as
far as interaction with acoustic waves is concerned) might be larger
than the observed penumbral area.

\section{SCATTERING OF SOUND WAVES BY SUNSPOTS}

When a wave is incident on an obstacle in its path, in addition to the
undisturbed wave, there is a scattered wave which spreads out from the
scatterer in all directions and which distorts and interferes with the
original wave. If the obstacle is very large compared with the
wavelength, i.e. in the limiting case of geometrical acoustics, the
scattered wave interferes with the undisturbed wave in such a way as
to create a sharp-edged shadow behind the object.

However if the size of the obstacle is comparable with the wavelength,
a variety of diffraction phenomena occur.

Here our scattering body is the sunspot, which we model as a patch on
the solar surface whose effect is to induce a fixed phaseshift
$\delta$ upon all incident rays.

Notwithstanding the fact that the asymptotic ray description breaks
down at the upper turning point and that the linearised adiabatic wave
equation is not even valid there, let us consider a ray originating at a
point A as shown in Fig.\ref{d}.

{\small
 \begin{figure}[h]
   \begin{center}
   \leavevmode
   \centerline{\epsfig{file=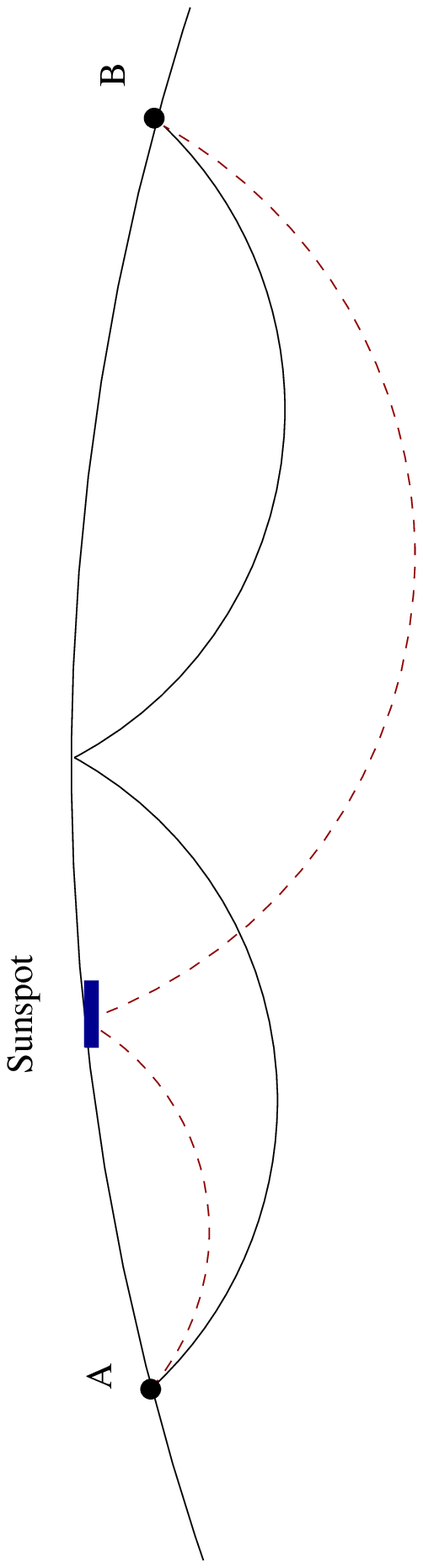,
               width=2.0cm,angle=270}}
   \end{center}
 \caption{\em Scattering of acoustic rays at sunspot.}
 \label{d}
 \end{figure}
  }

The wavefield $\psi$ observed at point B due to all singly reflected
rays from a source at point A can be represented by the undisturbed
reflected ray (continuous curve) plus the ray that is scattered off
the sunspot (dashed curve):

\begin{equation}
\psi (B)\propto 1+a {\rm e}^{i\delta}{\rm e}^{i\omega (\tau_{\rm r}-\tau_{\rm s})},\label{eq1}
\end{equation}
where $\omega$ is the angular frequency of the acoustic waves,
$\tau_{\rm r}$ is the wave travel time of the reflected ray and
$\tau_{\rm s}$ that of the scattered wave. The amplitude $a$ is
essentially the scattering strength of the sunspot, which depends on
the physical properties of the sunspots as well as the exact shape of
the `trapping potential', $k(r)=(\omega^2-\omega_{\rm c}^2)/c^2$
($\omega_{\rm c}$ is the acoustic cut-off frequency) at the upper
turning point. We omitted here the $1/r^2$ dependence of the
amplitudes because the lengths of the paths of the reflected and
scattered ray are nearly equal (identical for plane-parallel polytrope).

Since the waves propagate nearly vertically at the upper turning
point, we are neglecting any angular dependence of the scattered
wave. Also we are ignoring any possible angular variation in the
amplitude of the incident wave field.

In the Born approximation to scattering, this scattering strength is
proportional to the area of the sunspot. Thus, by observing sunspots of
different sizes, one might be able to constrain $a$.

The time delay caused by a sunspot as a function of the offset from
 the spot of the origin of the circle over which the signal is
 averaged (see Fig.\ref{a}) has been measured by Duvall (1995), and is
 shown in Fig.\ref{b}.

Fig.\ref{b} can be reproduced by averaging equation (\ref{eq1}) over
all azimuthal angles for a particular offset of the sunspot from the
centre of the observation ring and the result is shown by the dotted
line in Fig.\ref{b}.

It is important to notice that different sets of pairs of scattering
strengths, $a$, and phase shifts, $\delta$, can yield the same time
delays in Fig.\ref{b}. In Fig.\ref{g} we have plotted those scattering
strengths versus those phase shifts that are compatible with the measurement
of Fig.\ref{b}. 

{\small
 \begin{figure}[h]
   \begin{center}
   \leavevmode
   \centerline{\epsfig{file=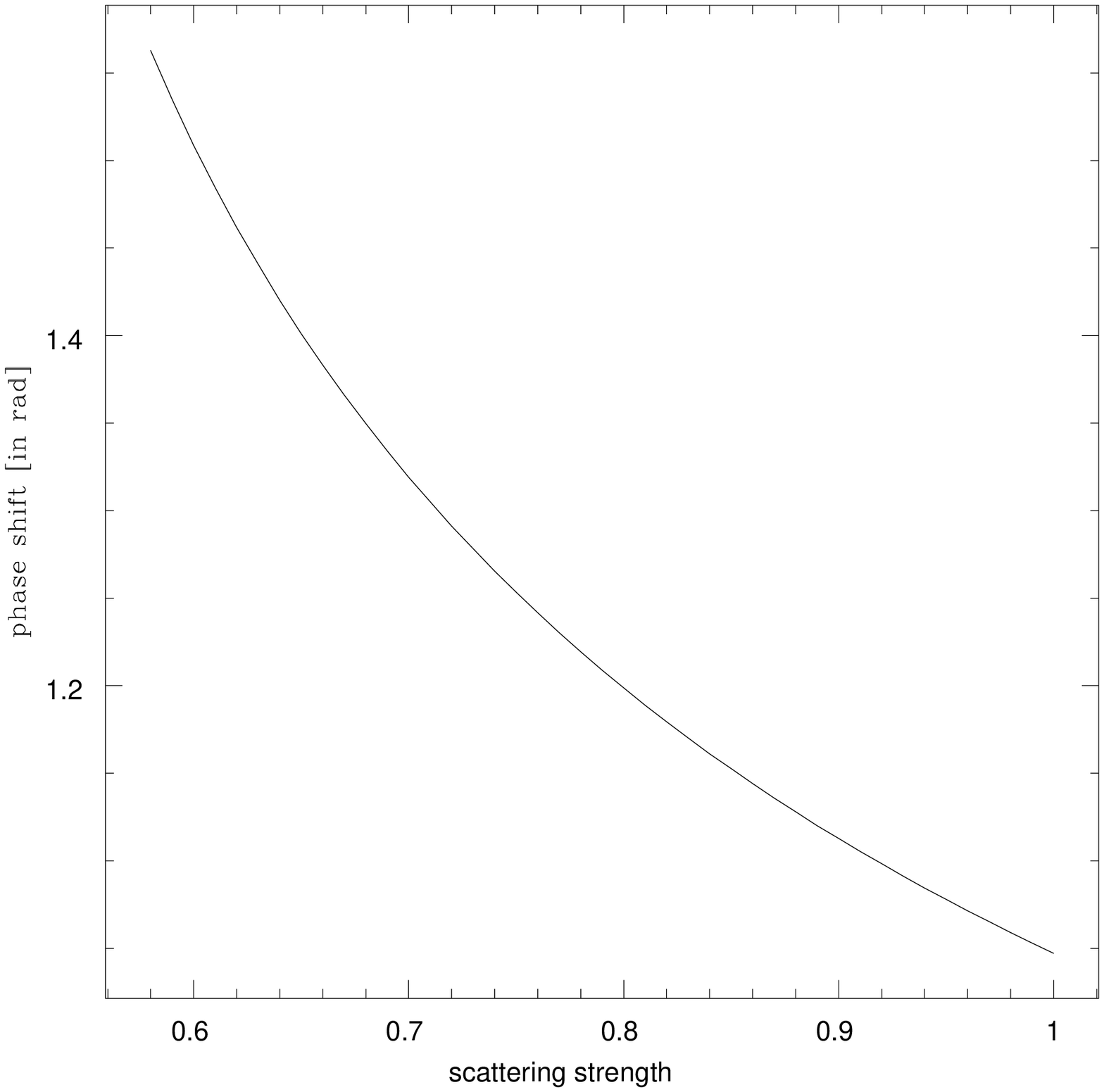,
               width=8.0cm}}
   \end{center} \caption{\em Scattering strength versus phase shift compatible with the observation by Duvall (1995).}  \label{g} \end{figure}
  }

We note that in order to produce a time delay of 0.56 minutes the
scattering strength has to lie between 0.6 and 1.0, causing the phase
shift induced by the sunspot to lie between 1.05 and 1.55 rad.

\section{DISCUSSION}

 As mentioned in the previous section, it is not possible to evaluate
the phase shift induced by a sunspot from the observed time delay
alone, unless we know the scattering strength of the sunspot. However,
a simple model suggests that this phase shift lies between 1.05 and
1.55 radians.  The correction to the value obtained for the
contribution from sunspots to the frequency shifts depends on this
phase shift. Moreover, it depends on the effective area of the
sunspots (i.e., the area of interaction with the acoustic waves),
which might be larger than that observed. Therefore, using the minimum
value for the phase shift obtained from the observed time delays, we
conclude that the contribution to frequency shifts coming from
sunspots is greater than one fifth. It might be significantly greater
if the effective area of the sunspots is larger than that observed, or
if the phase shifts induced by the sunspots is larger than the minimum
value required to reproduce Duvall's measurements.

\section{CONCLUSION}

We have shown how the changes of the frequencies of acoustic modes
during the solar cycle can be related to the time delays induced by
sunspots.

We have demonstrated that the time delays found in the presence of a sunspot
cannot be converted unambiguously into a phase shift due to a sunspot
without the knowledge of its scattering strength. We
can, however, set limits on the maximum and minimum values of 
the phase shift, those being 1.05 and 1.55 radians, respectively.

Accordingly, at the present, we can give only a lower limit to the
contribution from the sunspots to the frequency variations of global
modes; this lower limit is about one fifth of the observed
frequency shifts.  Moreover, by relating the frequency shifts to the
time delays as determined by the time-distance analysis of sunspots,
we predict the frequency dependence of these time delays to follow
equation (\ref{eq:dtau}).

We hope that in the near future time-distance observations will be
able to test the latter prediction, as well as to provide sufficient
data on sunspots of different sizes in order to allow for the
determination of the scattering strength of the sunspots as a function
of their effective area. This will yield a more accurate estimate of
the contribution of sunspots to the frequency shifts of global modes.

\section*{ACKNOWLEDGMENTS}

MC is very grateful to the Centro de Astrofisica da Universidade do Porto
for the use of the computer facilities. MC is supported by JNICT
(Portugal) through a grant BD/5519/95, PRAXIS XXI programme.

\end{document}